\documentclass{article}
\def\Ref#1{(\ref{#1})}
\usepackage{amsmath}
\usepackage{amssymb}
\usepackage{cite}
\def\d{{\rm d}}
\begin{document}
\begin{titlepage}
\noindent{\large\textbf{Static and dynamic phase transition in
multidimensional voting models on continua}}

\vskip 2 cm

\begin{center}{F.~Roshani$^a${\footnote
{farinaz@iasbs.ac.ir}}, A.~Aghamohammadi$^b${\footnote
{mohamadi@alzahra.ac.ir}}, \& M.~Khorrami$^b${\footnote
{mamwad@mailaps.org}}} \vskip 5 mm

\textit{ $^a$ Institute for Advanced Studies in Basic Sciences,
             P.~O.~Box 159,\\ Zanjan 45195, Iran. }

\textit{ $^b$ Department of Physics, Alzahra University,
             Tehran 19938-91167, Iran. }
\end{center}

\begin{abstract}
\noindent A voting model (or a generalization of the Glauber model
at zero temperature) on a multidimensional lattice is defined as a
system composed of a lattice each site of which is either empty or
occupied by a single particle. The reactions of the system are
such that two adjacent sites, one empty the other occupied, may
evolve to a state where both of these sites are either empty or
occupied. The continuum version of this model in a $D$-dimensional
region with boundary is studied, and two general behaviors of such
systems are investigated. The stationary behavior of the system,
and the dominant way of the relaxation of the system toward its
stationary state. Based on the first behavior, the static phase
transition (discontinuous changes in the stationary profiles of
the system) is studied. Based on the second behavior, the
dynamical phase transition (discontinuous changes in the
relaxation-times of the system) is studied. It is shown that the
static phase transition is induced by the bulk reactions only,
while the dynamical phase transition is a result of both bulk
reactions and boundary conditions.
\end{abstract}
{\textbf{PACS numbers:}} 64.60.-i, 05.40.-a, 02.50.Ga \\
{\textbf{Keywords:}} voting model, reaction-diffusion, phase
transition
\end{titlepage}
\section{Introduction}
The study of the reaction-diffusion systems, has an attractive
area. A reaction-diffusion system consists of a collection of
particles (of one or several species) moving and interacting with
each other with specific probabilities (or rates in the case of
continuous time variable). In the so called exclusion processes,
any site of the lattice the particles move on, is either vacant or
occupied by one particle. The aim of studying such systems, is of
course to calculate the time evolution of such systems. But to
find the complete time evolution of a reaction-diffusion system,
is generally a very difficult (if not impossible) task.

Various methods have been used to study the reaction-diffusion
system: analytical techniques, approximation methods, and
simulation. The success of the approximation methods, may be
different in different dimensions, as for example the mean field
techniques, working good for high dimensions, generally do not
give correct results for low dimensional systems. A large fraction
of analytical studies, belong to low-dimensional (specially
one-dimensional) systems, as solving low-dimensional systems
should in principle be easier
\cite{ScR,ADHR,KPWH,HS1,PCG,HOS1,HOS2,AL,RK2,AKK2,AM1,RK5}.

Various classes of reaction-diffusion systems are called
exactly-solvable, in different senses. In \cite{AA,RK3,RK4},
integrability means that the $N$-particle conditional
probabilities' S-matrix is factorized into a product of 2-particle
S-matrices. This is related to the fact that for systems solvable
in this sense, there are a large number of conserved quantities.
In \cite{BDb,BDb1,BDb2,BDb3,Mb,HH,AKA,MB,AAK}, solvability means
closedness of the evolution equation of the empty intervals (or
their generalization).

Consider a reaction-diffusion system (on a lattice) with open
boundaries. By open boundaries, it is meant that in addition to
the reactions in the bulk of the lattice, particles at the
boundaries do react with some external source. A question is to
find the possible phase transitions of the system. By phase
transition, it is meant a discontinuity in some behavior of the
system with respect to its parameters. Such discontinuities, may
arise in two general categories: in the stationary (large time)
profiles of the system, and in the time constants determining the
evolution of the system. In the first case, static phase
transitions are dealt with; in the second case, dynamical phase
transitions. For a review on dynamical phase transitions, one can
see for example \cite{HHA}.

There are systems for them the equation of motion of the one-point
function (the probability that a certain site be occupied) is
closed, that is independent of the more-point functions
\cite{GS,AAMS,SAK}. Among these systems is the so called voting
model (or a generalization of the Glauber model at zero
temperature). In \cite{MA1} a voting system on a one-dimensional
lattice was studied, for which at the boundaries of the lattice
there are injection or extraction of the particles. Based on the
evolution of the one-point functions, it was shown there that the
system exhibits two kinds of phase transitions: a static phase
transition, corresponding to a discontinuous change in the
stationary profile of the one-point function; and a dynamical one,
corresponding to a discontinuous change in the behavior of the
relaxation time of the system toward its stationary state. In
\cite{AM2,MAM,MA2,MA3,AM3}, the phase structures of extensions of
such systems on a one-dimensional lattice were investigated. All
of these are restricted to the case of a one-dimensional lattice.

Extending these investigations to higher-dimensional cases would
be interesting. Here, we want to study a multi-dimensional
extension of the voting model, on a continuum rather than a
lattice.

The scheme of the present article is as follows. In section 2, the
multidimensional voting model on continuum is presented, and the
evolution equation for the density of the particles is obtained.
In section 3, the time-independent of the system is studied and it
is shown that system exhibits a static phase transition. In
section 4, the relaxation of the system toward its stationary
state is studied and it is shown that the system exhibits a
dynamical phase transition. Section 5 is devoted to the concluding
remarks.
\section{Multidimensional voting models on continua}
In \cite{AM1} and \cite{MA1}, a one dimensional voting model (or a
generalization of the Glauber model at zero temperature) on a
lattice was defined as follows. Let the system consist of a
one-dimensional lattice, each of the sites of which are either
empty ($\emptyset$) or containing a single particle ($A$), and let
there be a reaction between two neighboring sites like
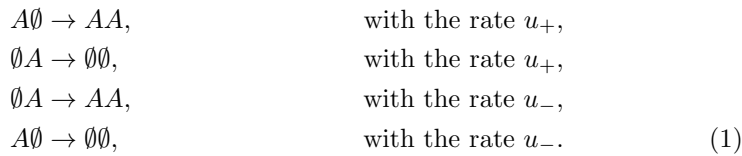
\begin{align}\label{1}
&A\emptyset\to AA,\qquad &\hbox{with the rate }u_+,\nonumber\\
&\emptyset A\to\emptyset\emptyset,\qquad &\hbox{with the rate }u_+,\nonumber\\
&\emptyset A\to AA,\qquad &\hbox{with the rate }u_-,\nonumber\\
&A\emptyset\to\emptyset\emptyset,\qquad &\hbox{with the rate }u_-.
\end{align}
In \cite{AM1}, an open lattice was investigated while in
\cite{MA1}, a lattice was studied at the boundaries of which
injection and extraction of particles could occur. It was shown
that these models are autonomous, meaning that the evolution
equation of the $n$-point functions contain only $n$- or
less-point functions. In fact, as it was seen in \cite{MA1},
\begin{equation}\label{2}
\frac{\d}{\d t}\,\langle n_i\rangle = u_+\,\langle n_{i-1}\rangle
+u_-\,\langle n_{i+1}\rangle-(u_+ +u_-)\,\langle n_i\rangle.
\end{equation}
Here $n_i$ is the particle number operator at the site $i$ of the
lattice.

Now consider a multi-dimensional lattice, each site of which is
either empty or occupied by a single particle, and let there be a
reaction like
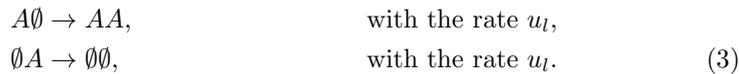
\begin{align}\label{3}
&A\emptyset\to AA,\qquad &\hbox{with the rate }u_l,\nonumber\\
&\emptyset A\to\emptyset\emptyset,\qquad &\hbox{with the rate
}u_l.
\end{align}
Here, we are considering the reaction between a site $i$ (the
right site), which is the ending point of the link $l$, and
another site (the left site) which is the starting point of the
same link. In a one-dimensional lattice, each site is the ending
point of two links, which had been denoted by $+$ and $-$. From
\Ref{3}, it is seen that the evolution equation for the one-point
function is
\begin{equation}\label{4}
\frac{\d}{\d t}\,\langle n_i\rangle =\sum_l [u_l\,(\langle
n_{i-l}\rangle-\langle n_{i-l}\,n_i\rangle)-u_l\,(\langle
n_i\rangle-\langle n_{i-l}\,n_i\rangle)],
\end{equation}
where by the site index $i-l$, it is meant a site which is the
starting point of the link $l$, the ending point of which is the
site $i$. It is seen that the two-point function in the right-hand
side of \Ref{4} cancel each other. So,
\begin{equation}\label{5}
\frac{\d}{\d t}\,\langle n_i\rangle =\sum_l u_l\,(\langle
n_{i-l}\rangle-\langle n_i\rangle).
\end{equation}

Now, assume that the one-point function is a slowly-varying
function of its argument ($i$). In this case, one can define a
smooth particle density function of the continuous position
variable $\mathbf{r}$, with
\begin{equation}\label{6}
\rho(\mathbf{r}_i):=\frac{1}{\mathcal V}\,\langle n_i\rangle,
\end{equation}
where $\mathbf{r}_i$ is position of the lattice site $i$, and
${\mathcal V}$ is the \textit{specific hypervolume} of a site.
Then \Ref{5} can be rewritten as
\begin{equation}\label{7}
\frac{\partial}{\partial t}\,\rho =\sum_l
u_l\,\left[-\boldsymbol{\delta}_l\cdot\nabla+\frac{1}{2}\,
(\boldsymbol{\delta}_l\cdot\nabla)^2\right]\,\rho,
\end{equation}
where $\boldsymbol{\delta}_l$ is the link-vector, equal to the
position of the ending point of the link $l$ minus the position of
the starting point of the link $l$, and higher-derivative terms
have been neglected. Using suitable coordinates for $\mathbf{r}$,
one can write the second derivative as
\begin{equation}\label{8}
\frac{1}{2}\,\sum_l u_l\,(\boldsymbol{\delta}_l\cdot\nabla)^2=
\sum_a\left(\frac{\partial}{\partial x^a}\right)^2,
\end{equation}
where $x^a$'s are the coordinates of $\mathbf{r}$. So, \Ref{7} is
rewritten as
\begin{equation}\label{9}
\frac{\partial}{\partial t}\,\rho
=(-\mathbf{v}\cdot\nabla+\,\nabla^2) \rho,
\end{equation}
where
\begin{equation}\label{10}
\mathbf{v}:=\sum_l u_l\,\boldsymbol{\delta}_l.
\end{equation}
Eq. \Ref{9} is nothing but a diffusion equation combined with a
drift velocity $\mathbf{v}$.

Suppose that \Ref{9} holds for the interior of the region $V$.
Integrating \Ref{9} on $V$, one arrives at
\begin{equation}\label{11}
\frac{\d}{\d t}\int_V\d V\;\rho=-\oint_{\partial V}\d S\;
\mathbf{n}\cdot\mathbf{v}\,\rho+\oint_{\partial V}\d S\;
\mathbf{n}\cdot\nabla\rho.
\end{equation}
The first term in the right-hand side is the rate of change of the
total number of the particles inside, as a consequence of the
drift, while the second term is the effect of injecting or
extracting particle at the boundary. The boundary condition
\begin{equation}\label{12}
\mathbf{n}\cdot\nabla\rho=\alpha-\beta\,\rho,\qquad \hbox{at the
boundary}
\end{equation}
corresponds to an injection rate of $\alpha$ per unit hyperarea of
the boundary, and an extraction rate of $\beta$ per unit hyperarea
per particle density at the boundary. In general, one can take
$\alpha$ and $\beta$ position-dependent.

Comparing \Ref{9} and \Ref{12} with eq. (7) of \cite{MA1}, it is
seen that one can transform eq. (7) of \cite {MA1} to \Ref{9} and
\Ref{12} through
\begin{align}\label{13}
\delta&=\sqrt{\frac{2}{u+v}},\nonumber\\
\mathbf{v}\cdot\hat{\mathbf{x}}&=\delta\,(u-v),\nonumber\\
\alpha^-&=\frac{a}{\delta\,u\,{\mathcal V}},\nonumber\\
\beta^-&=\frac{a+a'}{\delta\,u},\nonumber\\
\alpha^+&=\frac{b}{\delta\,v\,{\mathcal V}},\nonumber\\
\beta^+&=\frac{b+b'}{\delta\,v},
\end{align}
where the right-hand sides are the quantities defined in
\cite{MA1}, and the superscripts $-$ and $+$ refer to the left-
and right-boundaries, respectively.

From now on, for simplicity we restrict ourselves to the case that
the volume $V$ is a $D$-dimensional hyperball with radius $R$, the
boundary of which is a hypersphere.
\section{The time-independent state and the static phase transition}
Let $\rho_0$ be the time-independent solution to \Ref{9} and
\Ref{12}. Using the ansatz
\begin{equation}\label{14}
F_{\mathbf{q}}(\mathbf{r}):=\exp(\mathbf{q}\cdot\mathbf{r})
\end{equation}
(with $\mathbf{q}$ a constant vector) as a time-independent
solution to \ref{9}, one arrives at
\begin{equation}\label{15}
\mathbf{q}\cdot\mathbf{q}-\mathbf{v}\cdot\mathbf{q}=0,
\end{equation}
which leads to
\begin{equation}\label{16}
\mathbf{q}=\frac{1}{2}\,(\mathbf{v}+\mathbf{v}'),
\end{equation}
where $\mathbf{v}'$ is an arbitrary constant vector subject to the
condition
\begin{equation}\label{17}
\mathbf{v}'\cdot\mathbf{v}'=\mathbf{v}\cdot\mathbf{v}.
\end{equation}
So, one can write the general time-independent solution to \Ref{9}
as
\begin{align}\label{18}
\rho_0({\bf r})=&\int\d\Omega'\;\tilde
A(\Omega')\,F_{\mathbf{q}}(\mathbf{r}),\nonumber\\
=&\int\d\Omega'\;A(\Omega')\,
\exp\left\{\frac{1}{2}\,[(\mathbf{v}+\mathbf{v}')\cdot\mathbf{r}-
|\mathbf{v}+\mathbf{v}'|\,R]\right\},\nonumber\\
=:&\int\d\Omega'\;A(\Omega')\, \exp[G(\mathbf{v}',\mathbf{r})],
\end{align}
where $\Omega'$ denotes the angular coordinates of $\mathbf{v}'$,
and $A$ is an arbitrary function. It is easy to see that the
maximum value of $G$ is zero, and this maximum value is reached at
a point on the boundary ($r=R$), where $\mathbf{r}$ is parallel
with $\mathbf{v}+\mathbf{v}'$.

For large values of $R$ and $r$, $G$ is a rapidly-varying function
and the integral is mainly determined from that point of the
integration region which maximizes $G$. Generally, there may be
two such points: One point is
\begin{equation}\label{19}
\mathbf{v}'_1=-\mathbf{v}.
\end{equation}
The other point is
\begin{equation}\label{20}
(\mathbf{v}'_2+\mathbf{v})\cdot\mathbf{r}=|\mathbf{v}'_2+\mathbf{v}|\,r,
\qquad\hbox{for $r=R$},
\end{equation}
which means that $\mathbf{q}$ is parallel with $\mathbf{r}$. As
the angle between $\mathbf{q}$ and $\mathbf{v}$ cannot exceed
$\pi/2$, the second point exists only if the angle between
$\mathbf{r}$ and $\mathbf{v}$ is less than $\pi/2$. One has
\begin{equation}\label{21}
G(\mathbf{v}'_1,\mathbf{r})=0,
\end{equation}
and
\begin{align}\label{22}
G[\mathbf{v}'_2(\mathbf{r}),\mathbf{r}]=&
G[\mathbf{v}'_2(R\,\mathbf{r}/r),\mathbf{r}]+O[(R-r)^2],\nonumber\\
=&\frac{(r-R)\,|\mathbf{v}+\mathbf{v'}_2|}{2}+O[(r-R)^2],\nonumber\\
=&\frac{(r-R)\,\mathbf{v}\cdot\mathbf{r}}{R}+O[(r-R)^2].
\end{align}
Using \Ref{21} and \Ref{22}, one arrives at
\begin{equation}\label{23}
\rho_0(\mathbf{r})\sim\begin{cases} C_1(\Omega),& r\sim R,
\mathbf{r}\cdot\mathbf{v}<0,\\
C_1(\Omega)+C_2(\Omega)\,\exp[\frac{(r-R)\,\mathbf{v}\cdot\mathbf{r}}{R}],&
r\sim R, \mathbf{r}\cdot\mathbf{v}>0,
\end{cases}.
\end{equation}
From this,
\begin{equation}\label{24}
\nabla\rho_0(r=R)\propto\mathbf{n}\,(\mathbf{n}\cdot\mathbf{v})\,
\theta(\mathbf{n}\cdot\mathbf{v}),\qquad R\to\infty,
\end{equation}
Where $\theta$ is the step function. It is seen that in the
thermodynamic limit ($R\to\infty$), the density profile at the
boundary is stationary, unless $\mathbf{v}\cdot\mathbf{r}>0.$ So,
changing $\mathbf{v}$ one can induce a discontinuous change in the
slope of the profile density at the boundary. This is the static
phase transition, which is seen to be independent of the injection
and extraction terms, but dependent on the drift velocity.
\section{The relaxation of the system toward the stationary state,
and the dynamic phase transition} Starting from \Ref{9} and
\Ref{12}, one arrives at
\begin{align}\label{25}
\frac{\partial}{\partial t}\,(\rho-\rho_0)
&=(-\mathbf{v}\cdot\nabla+\,\nabla^2) (\rho-\rho_0),\nonumber\\
&=:h\,(\rho-\rho_0),
\end{align}
and
\begin{equation}\label{26}
\mathbf{n}\cdot\nabla(\rho-\rho_0)=-\beta\,(\rho-\rho_0),\qquad
\hbox{at the boundary}
\end{equation}
where $\rho_0$ is the time-independent solution to \Ref{9} and
\Ref{12}. Let $\psi$ be an eigenfunction of $h$ corresponding to
the eigenvalue $E$. Using the ansatz \Ref{14} in the eigenvalue
equation corresponding to $h$, one arrives at
\begin{equation}\label{27}
\mathbf{q}\cdot\mathbf{q}-\mathbf{v}\cdot\mathbf{q}=E,
\end{equation}
which leads to
\begin{equation}\label{28}
\mathbf{q}=\frac{1}{2}\,(\mathbf{v}+\mathbf{v}'),
\end{equation}
where $\mathbf{v}'$ is an arbitrary constant vector subject to the
condition
\begin{equation}\label{29}
\mathbf{v}'\cdot\mathbf{v}'=\mathbf{v}\cdot\mathbf{v}+4\,E.
\end{equation}
So, one has
\begin{equation}\label{30}
\psi(\mathbf{r})=\exp(\mathbf{v}\cdot\mathbf{r}/2)\int\d\Omega'
\;A(\Omega')\,\exp(\mathbf{v}'\cdot\mathbf{r}/2),
\end{equation}
where $A$ is to be found so that the boundary condition \Ref{26}
is satisfied with $\psi$.

If the righthand side of \Ref{29} is positive, then $\mathbf{v}'$
is real and for large $r$, one can approximate $\psi$ like
\begin{equation}\label{31}
\psi(\mathbf{r})\sim\exp(\mathbf{v}\cdot\mathbf{r}/2)
\,A(\Omega)\,\exp(v'\,r/2),
\end{equation}
where $\Omega$ is the angular coordinates corresponding to
$\mathbf{r}$. The boundary condition \Ref{16}, then becomes
\begin{equation}\label{32}
\left[\frac{v'}{2}+\beta(\Omega)+\frac{\mathbf{n}\cdot\mathbf{v}}{2}
\right]\,A(\Omega)=0.
\end{equation}
This has a nonzero solution for $A$, provided the parenthesis
vanishes for some $\Omega$. As $\beta\geq 0$, this happens for
some (real) positive $v'$, if and only if
\begin{equation}\label{33}
\min\left[\beta(\Omega)+\frac{v\,\cos\phi}{2}\right]<0,
\end{equation}
where $\phi$ is the angle between $\mathbf{r}$ and $\mathbf{v}$.
If \Ref{33} holds, then the range of $v'$ for which a nonzero
solution to \Ref{32} for $A$ exists is
\begin{equation}\label{34}
0\leq v'\leq
-\min\left[\beta(\Omega)+\frac{v\,\cos\phi}{2}\right].
\end{equation}
(This is true for more-than-one dimensional space. If the space is
one-dimensional, $v'$ has only one acceptable value, as the
parenthesis in \Ref{32} has only two values at most one of them
can be zero.)

If \Ref{33} holds, then there exists eigenvalues $E$ for $h$, with
$E>-\mathbf{v}\cdot\mathbf{v}/4$. Otherwise, all of the
eigenvalues of $h$ are less than or equal to
$-\mathbf{v}\cdot\mathbf{v}/4$. The relaxation time of the system
is
\begin{equation}\label{35}
\tau=-\frac{1}{E_{\mathrm{max}}},
\end{equation}
where $E_{\mathrm{max}}$ is the largest eigenvalue of $h$. The
largest value of $E$ is either $-\mathbf{v}\cdot\mathbf{v}/4$, or
the value obtained from \Ref{29} for the largest value of $v'$.
So,
\begin{equation}\label{36}
\tau=\begin{cases}
     \frac{4}{\mathbf{v}\cdot\mathbf{v}},&
     \min\left[\beta(\Omega)+\frac{v\,\cos\phi}{2}\right]>0\\
     \frac{4}{\mathbf{v}\cdot\mathbf{v}-
     \{\min[2\beta(\Omega)+v\,\cos\phi]\}^2},&
     \min\left[\beta(\Omega)+\frac{v\,\cos\phi}{2}\right]<0
     \end{cases}
\end{equation}
In the first case, the system is in the fast dynamical phase, in
which the relaxation time does not depend on the boundary
condition. In the second case, the system is in the slow dynamical
phase, in which the relaxation time is larger and does depend on
the boundary conditions. This is the dynamical phase transition.
\section{Concluding remarks}
It was seen that the so-called voting model defined on a
one-dimensional lattice, has a natural analog on a
multidimensional continuum. It was seen that there are two kinds
of phase transition, a static one corresponding to a discontinuous
change in the behavior of the stationary profile of the system,
and a dynamical phase transition corresponding o the relaxation of
the system toward its stationary state. The static phase
transition is controlled by the bulk reactions, while the
dynamical phase transition is controlled by the bulk reactions and
the boundary conditions both. This is analogous to what seen for
the case of a one-dimensional lattice.

There are, however, differences. In the multidimensional case, the
static transition occurs when the direction of the drift velocity
is changed. This can happen without being necessary that the drift
velocity vanishes. In the one-dimensional case, however, the
static phase transition occurs only when the drift velocity passes
zero. The reason is that in one-dimension, the only way to change
the direction of a vector smoothly, is that the vector vanishes at
some point.

The second difference concerns the dynamical phase transition, to
be more precise, the largest eigenvalues of the operator $h$
defined in \Ref{25}. In the one-dimensional case and in the slow
phase, there is only one eigenvalue greater than the largest
eigenvalue corresponding to the fast phase. In the
multidimensional case, however, in the slow phase the spectrum of
$h$ contains a continuous region the lower bound of which is the
largest eigenvalue of $h$ in the fast phase. This means that in
the one-dimensional case and in the slow phase, there is a largest
relaxation time and a gap between this and the next largest
relaxation time, while in the multidimensional case, there is no
such gap.

\end{document}